\documentclass[twocolumn, showpacs]{revtex4}

\usepackage{amssymb,amsmath,graphics,graphicx} 

\begin{document}

\title{Nonlinear diffusion effects on biological population spatial patterns }

\author{Eduardo H. Colombo$^1$  and Celia Anteneodo$^{1,2}$}
\affiliation{$^1$Department of Physics, PUC-Rio, Rio de Janeiro, Brazil \\
$^2$National Institute of Science and Technology for Complex Systems, Rio de Janeiro, Brazil.}

\begin{abstract}
Motivated by the observation that anomalous diffusion 
is a realistic feature in the dynamics of biological populations, 
we investigate its implications  in a paradigmatic model  
for the evolution of a single species density $u(x,t)$.
The standard model includes growth and competition in a logistic expression,  and 
spreading is modeled through normal diffusion. Moreover, the competition 
term is  nonlocal, which has been shown to give rise to  spatial patterns.    
We generalize the diffusion term   
through the nonlinear form $\partial_t u( x,t) = D \partial_{xx} u(x,t)^\nu$ 
(with  $D, \,\nu>0$), encompassing the cases where
the state-dependent diffusion coefficient either 
increases ($\nu>1$) or decreases ($\nu<1$) with the density, yielding subdiffusion or 
superdiffusion, respectively.  
By means of numerical simulations and analytical considerations, we display 
how that nonlinearity alters the phase diagram.
The type of diffusion imposes critical values  of the model parameters for the onset of patterns 
and  strongly influences their shape, inducing fragmentation in the subdiffusive case. 
The detection of the main persistent mode allows analytical prediction of the  critical 
thresholds.

\end{abstract}


\maketitle

\section{Introduction}

Pattern formation in population dynamics has been studied both experimentally and 
theoretically.
In experiments,  the dynamics of insects and bacterial colonies, amongst others,  
have been observed \cite{woolley,fao,chaotic,experimental,nonlinear1}. 
From the theoretical viewpoint,  mean-field descriptions render  macroscopic 
or mesoscopic approximations to describe the behavior of such complex systems.  
To take into account spatial inhomogeneities, one may construct a partial differential 
equation that rules the temporal evolution of the population density $u(\vec{x},t)$, 
a function of the spatial position $\vec{x}$ and time $t$. 
Within this family of models, a standard one was first introduced by Fisher  \cite{fisher}. 
It consists of a reaction-diffusion  equation, taking into account 
growth and competition in the usual logistic form. 
Recently, a generalization of Fisher equation (FE) was introduced  \cite{fuentes,analytical,applicability}, namely, 
\begin{equation}  \label{g0fisher}
\frac{\partial}{\partial t}u( \vec{x},t) = D \nabla^2 u(\vec{x},t) + u(\vec{x},t)(a - bJ[u(\vec{x},t)]) \,,
\end{equation}
where  $D$, $a$, $b$ are  positive parameters and 
$J$ is a functional of the density embodying  nonlocality: 
\begin{equation}
J[u(\vec{x},t)] = \int_\Omega f(\vec{x},\vec{x}') u(\vec{x}',t) d\vec{x}' \,.
\end{equation}
In the particular case $f(\vec{x},\vec{x}')=\delta(\vec{x}-\vec{x}')$, 
the original (local) FE is recovered.
The introduction of the nonlocal form of the competition term is motivated by the 
consideration that  products released in the environment 
by the individuals  may either harm or support neighbors' growth. 
Interestingly, this nonlocal component was shown to give rise to 
the formation of steady spatial patterns \cite{fuentes}.
Diverse variants have been studied before. 
As influence functions $f(\vec{x},\vec{x}')$, square and smooth forms have been 
considered \cite{fuentes,analytical}. Nonlocality in the reproduction rate \cite{fao},   
dimensionality   \cite{brigatti2} and  fluctuation effects \cite{brigatti} have been  
investigated too.  
In all those cases, however, spatial spread was described by normal diffusion. 
Meanwhile, there are indications  that the 
spreading of biological populations is not due to purely random motion but 
influenced by the density, either to favor or to avoid crowding  \cite{murray,gurtin,gregarious}. 
Hence dispersal is guided by a state-dependent diffusion coefficient rather than by a constant one. 

An important class of generalized diffusion equations is constituted  by 
the porous media equation $\partial_t u = \partial_{xx} u^\nu$, originally defined for 
$\nu>1$ \cite{porous}. 
Although it was later extended to real $\nu>-1$ \cite{bukman}, here we will restrict our 
analysis to $\nu>0$. 
Nonlinear diffusion is ruled  by a state-dependent  diffusion coefficient,  
  proportional to $u^{\nu-1}$, hence embracing the cases where
the coefficient either grows \cite{sub} or decreases \cite{super} with the density $u$. 
The generalization of Arrhenius law \cite{escape}, the performance of 
thermal ratchets \cite{bmotors}, and other properties such as aging \cite{more} that arise 
under this kind of diffusion have been studied before. 
Nonlinear diffusion equations in higher dimensions \cite{dim} 
or even with space-fractional derivatives \cite{frac} have 
been analytically solved. 
The nonlinearity leads to anomalous diffusion \cite{bukman}:  
either superdiffusion for 
$\nu<1$ or subdiffusion for $\nu>1$, recovering normal diffusion when $\nu=1$.
Microscopically, high density regions can slow down ($\nu<1$) or intensify ($\nu>1$)
individual displacements, as a consequence of homophilic  
behaviors that rule the dynamics of self-diffusion favoring 
or not the mobility among other individuals.  
While $\nu<1$ reflects a reaction to sparseness (with high diffusion coefficient where 
the density is low), on the contrary, $\nu>1$ is associated to immobilization 
in poorly populated regions.

We will analyze the effects of nonlinear diffusion on pattern formation by 
 considering the one dimensional  generalized  FE   with nonlocal competition 
\begin{equation}  \label{gfisher}
\frac{\partial}{\partial t}u(x,t) = D\frac{\partial^2}{\partial x^2} u^\nu(x,t) + 
u(x,t)(a - bJ[u(x,t)])\,. 
\end{equation}
Since alternative forms of the functional $f(x,x')$ 
do not yielded substantially different results \cite{fuentes}, 
we will restrict our analysis  to the case where $f(x,x')$ 
is constant for $x-w\le x'\le x+w$, 
and zero otherwise, namely $f(x,x')=\frac{1}{2w}\Theta(w - |x-x'|)$, 
where $\Theta$ is the Heaviside step function. 

\section{Results}

Numerical integration of Eq.~(\ref{gfisher}) was performed 
by means of a standard forward-time centered-space scheme with integration time step 
$dt \le 10^{-3}$ and width of spatial grid cells  $dx\le 0.1$. 
We set periodic boundary conditions, with periodic domain size $L=100$. 
As initial condition we considered small amplitude random perturbations   
either  above the null state or around the nontrivial homogeneous solution. 
We also considered  square pulses (with small random fluctuations) 
with zero values everywhere else.

\begin{figure} [h!]
\centering
\includegraphics*[bb=70 320 500 750, width=0.95\columnwidth]{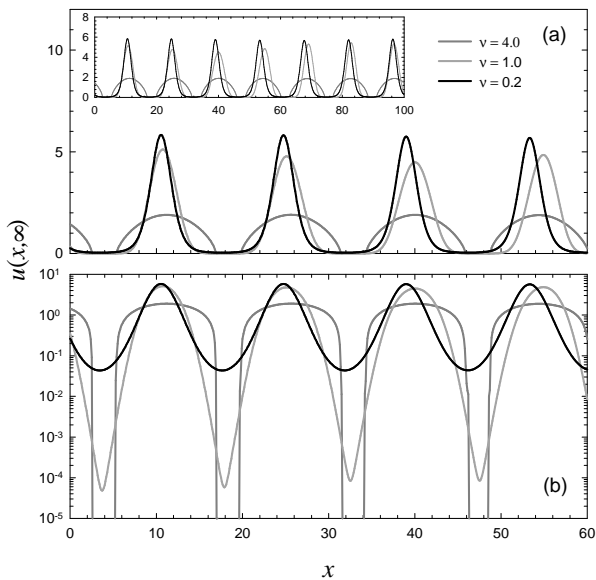}
\caption {Longtime patterns obtained from numerical integration 
of Eq. (\ref{gfisher}),  with  
$a=b=1$, $L=100$, $D=0.1$, $w=10$ and  different values of 
$\nu$ indicated on the figure, in (a) linear  and (b) logarithmic  scales. 
In the inset, the longterm profiles in the full grid are shown.
The profiles are plotted for $t=200$, but they remain unchanged after $t \simeq 100$.
} 
\label{fig:patterns}
\end{figure}

\begin{figure} [t!]
\centering
\includegraphics*[bb=70 320 500 750, width=0.95\columnwidth]{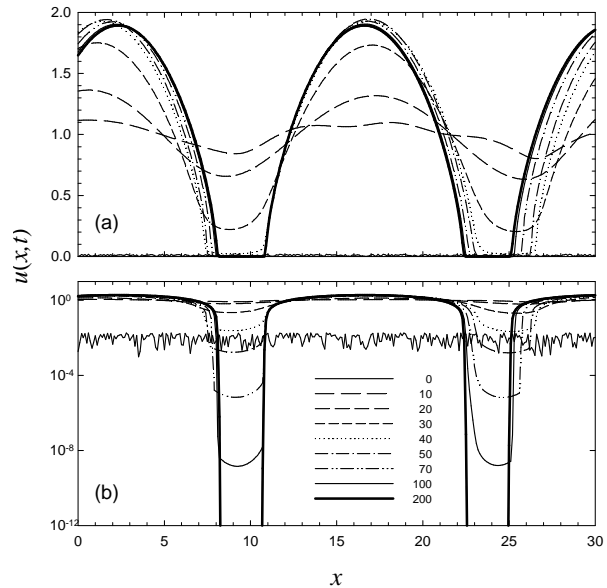}
\caption{Time evolution of the density profile obtained from numerical integration 
of Eq. (\ref{gfisher}) with  
$a=b=1$, $L=100$, $D=0.1$, $w=10$ and $\nu=4$, represented  for different times $t$ 
indicated on the figure, in (a) linear and (b) logarithmic scales. 
The thick line corresponds to $t=200$. 
} 
\label{fig:patterns_time}
\end{figure}
 
Typical longtime patterns, robust under changes in the initial conditions 
here considered,  are shown in Fig.~\ref{fig:patterns}. 
Notice that, while the number of peaks is not affected by changing $\nu$, 
the form of the patterns becomes substantially different. By increasing $\nu$, 
the width (inverse concavity) of the crests 
increases and the density at the valleys decreases, such that for $\nu>1$ disconnected 
regions can arise.

Figure~\ref{fig:patterns_time} shows the time evolution for $\nu=4$, 
starting with small random values of the density $u(x,0)$. It rapidly increases for all $x$ 
towards  the level corresponding to the homogeneous solution, $u_0=a/b$ ($t<10$), 
while patterns develop. 
After $t=100$ no substantial changes are detected at the crests. 
Between successive crests, the density tends zero (exponentially fast with time). 
This fragmentation or  
clusterization process \cite{clustering} yields isolated population groups  
(clusters). Therefore fluxes between clusters are eliminated in the long time limit. 
This phenomenon is crucial in connection with ``epidemic'' spreads within a population.

\subsection{Stability analysis}

\begin{figure} [t!]
\centering
\includegraphics*[bb=70 460 500 750, width=0.95\columnwidth]{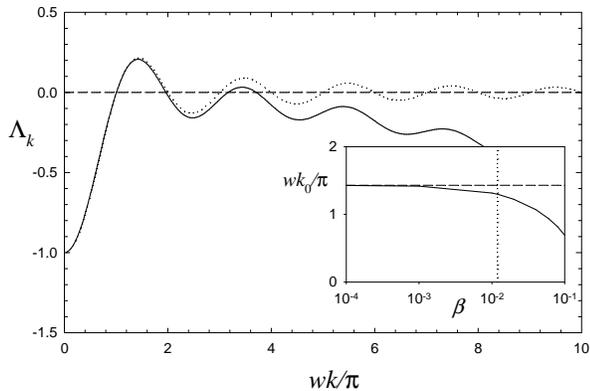}
\caption {Dispersion relation $\Lambda_k=\lambda_k/a$ vs scaled  $k$ (solid line), 
for $\beta=5\times 10^{-4}$. 
The dotted line corresponds to the term $-\sin(wk)/(wk)$ and the dashed line 
to the zero line, drawn for comparison. 
In the inset, we show the position of the first maximum $k_0$ as a function of 
$\beta\equiv  \nu D  u_0^{\nu-1}/(a w^2)$. The vertical dotted line indicates  
the instability threshold.
} 
\label{fig:lambda}
\end{figure}

To determine the stability conditions we follow the standard  
procedure of considering a first-order  
perturbation around the homogeneous solution $u_0=a/b$: 
\begin{equation}\label{perturbation}
u(x,t) = u_0 + \epsilon\exp( ikx  + \lambda_k t) \, ,
\end{equation}
where  $\epsilon$ is the perturbation initial amplitude, $k$ the wave number 
and $\lambda_k$ is the exponential rate of temporal behavior.   
Substituting Eq.~(\ref{perturbation}) into Eq.~(\ref{gfisher}) 
gives the dispersion relation 
\begin{equation}\label{dispersion}
\lambda_k = -\nu D u_0^{\nu-1}\,k^2 - a \frac{\sin(w k)}{w k}   \, ,
\end{equation}
that generalizes the one  obtained by Fuentes {\em et al.} \cite{analytical}.
Defining the nondimensional rate $\Lambda_k \equiv \frac{\lambda_k}{a}$, 
Eq.~(\ref{dispersion}) can be rewritten in a single parameter form as
\begin{equation}\label{dispersion2}
\Lambda_k =  -\beta (w k)^2 - \frac{\sin(w k)}{w k} \,,
\;\;\;\mbox{   with } \beta \equiv \frac{\nu D  \,u_0^{\nu-1}}{a\,w^2} \,.
\end{equation}
Negative  $\Lambda_k$ means relaxation back to the uniform state.  
Figure~\ref{fig:lambda} depicts the dispersion relation  in a 
typical case where $\Lambda_k$ can take positive values allowing  
instability growth. 

\begin{figure} [h!]
\centering
\includegraphics*[bb=70 460 500 750, width=0.95\columnwidth]{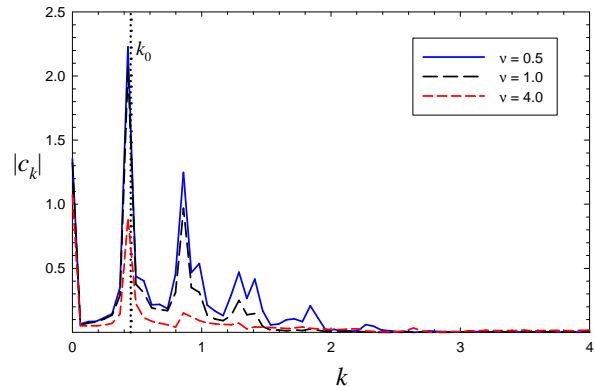}
\caption {(Color online) Fourier spectra for the longtime patterns shown in Fig.~\ref{fig:patterns}. 
The vertical dotted line indicates the position of the dominant mode $k_0$.
} 
\label{fig:fourier}
\end{figure}

Even if the analysis at short times does not guarantee 
the  later evolution towards a stationary state, the mode  
with largest growth rate $k^*$ (absolute maximum of the dispersion relation $\Lambda_k$ vs $k$) 
could play a crucial role. 
This mode  will excite other wavelengths though the coupling   
nonlinear term, however, if they are damped,  $k^*$ will  remain selected and its harmonics will 
shape the patterns. 
Substitution of the Fourier series expansion 
$u(x,t)=\sum_{k=-\infty}^\infty c_k(t) \exp(ikx)$  
 into Eq.~(\ref{gfisher}), when $\nu=1$, leads to the following evolution 
equations for the Fourier coefficients $c_k$  \cite{analytical}
\begin{equation}\label{fourier}
\frac{d c_k }{dt} = -Dk^2 c_k +a c_k -b \sum_m c_m \bar{c}_{k-m} \frac{\sin mw}{mw} \,.
\end{equation}
These equations are highly coupled through the last nonlinear term.  
If $\nu\neq 1$, there will be still an additional nonlinearity in the first term of the righthand side, 
anyway let us consider the case where the first term is very small, 
allowing the existence of unstable modes.  
The amplitude of the mode corresponding to the uniform state, $c_0$, grows with rate $a$ 
until stabilization, as observed in numerical simulations, 
e.g., in the example of Fig.~\ref{fig:patterns_time} the level $u_0=a/b=1$ is attained at 
times of order $1/a$.  
The mode with largest initial (positive) rate  quickly develops and 
keeps dominating at intermediate timescales.   
Notice in Fig.~\ref{fig:patterns_time} an almost perfect 
sinusoidal profile at time $t\simeq 20$. 
If a unique mode contributes 
to the sum in Eq.~(\ref{fourier}), it grows with rate given by Eq.~(\ref{dispersion})   
until stabilization while the remaining modes will be dumped. Actually a 
set of undamped harmonics, characteristic of each value of $\nu$, also persists to shape the profiles. 
Typical Fourier spectra for the long-term patterns are shown in   Fig.~\ref{fig:fourier}.
Although we do not have a rigorous mathematical proof, we will see that 
numerical results indicate that the dominant persistent mode, defining 
pattern wavelength, is the fastest growing one at short times.

$\Lambda_k$ possesses infinite local maxima located at $k_n$, $n=0,1,\ldots$. 
Since  the absolute  maximum is the first one, then $k^*=k_0$   (see Fig.~\ref{fig:lambda}). 
In the inset of Fig.~\ref{fig:lambda}, 
$k_0$ (numerically obtained) is plotted as a function of  $\beta$.  
For sufficiently small $\beta$, 
$\Lambda_k$ is dominated by the last term in Eq.~(\ref{dispersion2}),  
yielding $k_0 =\theta_0/w$ with $\theta_0 \simeq 1.43 \,\pi$. 
Fig. ~\ref{fig:fourier} that the dominant mode of long time patterns is in good accord with $k_0$.

Perturbations to the homogeneous solution  vanish if  $\Lambda_k < 0$. 
Since $\sin(wk)$ is bounded, then the instability condition $\Lambda_k>0$ implies  
\begin{equation} \label{cond1}
\beta  <  (w k)^{-3}    \,.  
\end{equation}
For the mode with largest growth,   considering the approximation 
$k_0 =\theta_0/w \simeq 1.43\pi/w$, one has the instability condition
\begin{equation} \label{cond2}
 \beta \equiv \frac{\nu D \,u_0^{\nu-1}}{a \,w^2} < \theta_0^{-3} \simeq (1.43\pi)^{-3} \,.
\end{equation} 
This is equivalent to requiring the positivity of the first maximum.    
Notice in the inset of Fig.~\ref{fig:lambda} that  $k_0  \simeq 1.43\pi/w$ remains  a good
approximation in the whole instability region, below the threshold (vertical dotted line in the inset). 
Beyond this point the maximum becomes negative, hence the homogeneous 
solution recovers its stability for any wavelength. 
Equation (\ref{cond2}) defines the critical value 
$\beta_c \equiv \theta_0^{-3}$ for the onset of patterns.

\begin{figure} [b!]
\centering
\includegraphics*[bb=70 320 500 750, width=0.95\columnwidth]{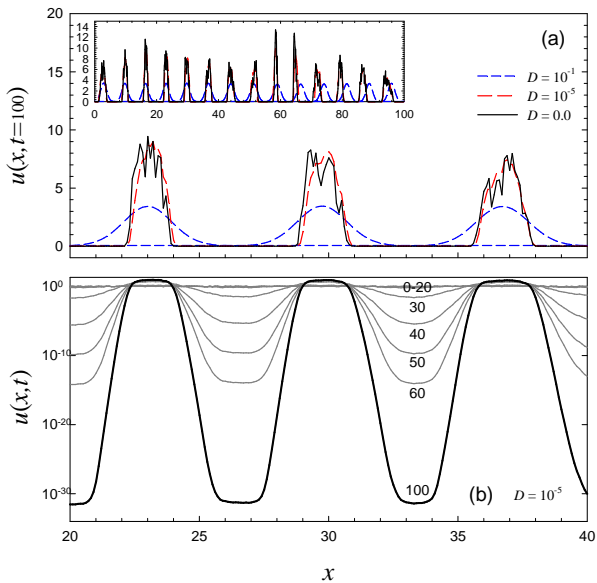}
\caption { (Color online)  
(a) Longtime patterns obtained from numerical integration 
of Eq. (\ref{gfisher}) up to time $t=100$,  with  
$a=b=1$, $\nu=1$, $w=5$, $L=100$  and  different values of 
$D$ indicated on the figure. 
In the inset, the profiles in the full grid are shown. 
In (b) the time evolution for $D=10^{-5}$ is shown in  logarithmic  scale. 
Times are indicated on the figure. 
} 
\label{fig:zerodif}
\end{figure}

As intuitively expected, on the basis of the homogenizing role of diffusion, 
the inequality in Eq.~(\ref{cond2}) indicates that 
the diffusion constant can not exceed a limiting value for the perturbation 
to depart from the homogeneous state. 
In accord with Eq. (\ref{cond2}), in the limit $D\to 0$, patterns are also observed.
Then, diffusion is not a necessary ingredient for the onset of patterns  
but has a role in pattern shaping.
Figure~\ref{fig:zerodif}(a) shows the density profiles that emerge 
for different values of $D$ in the normal 
case $\nu=1$. For $D=0$ patterns are noisy due to the lack of 
the smoothing effect of diffusion and the 
amplitude is less uniform but the wavelength $\ell$ is well defined. 
Moreover, between bumps, the density tends to zero as in the 
subdiffusive case of Fig. \ref{fig:patterns_time}. 
The width of the bump at zero height,  $2x_0$, is also well defined. 
Beyond fluctuations, the results are robust, 
at least under the types of initial conditions analyzed. 
In the absence of diffusion, the steady state must verify $u(x)[a-b J(x)]=0$, 
then either $u(x)$ vanishes or its integral 
within the interval $(x-w,x+w)$ must adopt the constant value $a/b$. 
The former case requires that the null solution becomes stable in some regions. 
The latter requires that  
each cluster does not see the neighbors and any point in the cluster must be influenced by the full cluster. 
This means that $2x_0 \le w \le \ell-2x_0$, which is verified in numerical experiments.

For $D=10^{-5}$, patterns are still noisy at $t=100$, but are expected to smooth out at much longer times. 
In this case also the density between crests goes to zero exponentially fast with time,  
as can be seen in Fig.~\ref{fig:zerodif}(b). 
In the case of the figure, the clusterization occurs for $D \lesssim 10^{-3}$ while for larger values 
of $D$ not only the crests but 
also the valleys stabilize in a finite value as time goes by. 
Then clusterization occurs for $D$ below a threshold value only.  
It is noteworthy the resemblance of  the noisy profiles with those observed 
in experiments with bacteria \cite{experimental}. But here  clusters arise naturally, without 
imposing absorbing nor zero flux boundaries.

If $D\neq0$, Eq.~(\ref{cond2}) predicts the existence of a minimal value of the interaction 
range $w$ required for pattern 
formation, with all other parameters kept fixed.  
This critical value does depend on the kind of diffusion 
through the factors $\nu$ and $u_0^{\nu-1}$.
Notice also that for $\nu\neq 1$ there is influence of $u_0$ too, which is absent in the normal case ($\nu=1$).

According to the hypothesis that $k_0$ is the characteristic wavenumber of the stationary pattern, 
and taking into account that boundary conditions are periodic (i.e., an integer number of 
wavelengths must accommodate into the size of the system $L$), 
the number of maxima $m$ is given (in average) by
\begin{equation} \label{aprox}
m = \frac{k_0L }{2\pi}=\frac{\theta_0}{2\pi} \frac{L}{w} \simeq 0.715\frac{L}{w}  \,. 
\end{equation}
Even in the cases when Eq.~(\ref{aprox}) gives an integer value, 
it is expected to furnish the number of peaks observed in the average. 
In practice, depending on the initial conditions,  the  crests 
come out and grow accommodating its number approximately to the rounded value of $m$. Fluctuations in 
the effective $m$ are larger the larger  $m$  or the further is $m$ from an integer value. 
For instance in the example of Fig. \ref{fig:patterns}, instead of seven, eight crests 
are observed in some realizations, being $m \simeq 7.15$.

\begin{figure} [b!]
\centering
\includegraphics*[bb=70 460 500 750, width=0.95\columnwidth]{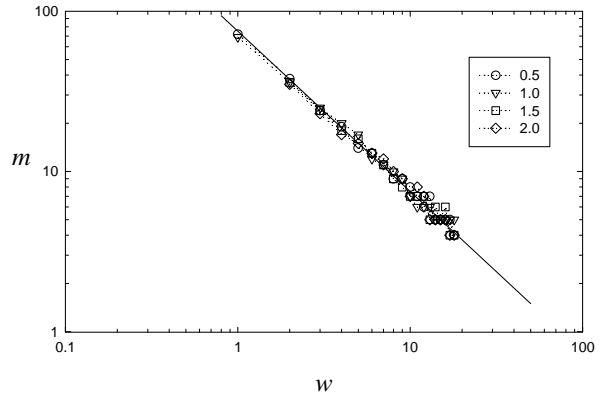}
\caption {Number of maxima $m$ as a function of the nonlocality width $w$, for 
$a=b=1$, $L=100$ and $D=0.01$, and different values of $\nu$ indicated on the figure. 
The solid line corresponds to the approximate 
theoretical relation Eq.~(\ref{aprox})
 and the symbols to the outcomes of numerical simulations. 
} 
\label{fig:nmax}
\end{figure}

These observations can be verified through numerical integration of the evolution 
equation. 
In Fig.~\ref{fig:nmax}, we show the  number of maxima $m$ of the patterns as a function  
of $w$ together with the theoretical prediction given by Eq. (\ref{aprox}).
An excellent agreement between 
theoretical  and numerical outcomes can be observed.  
Then, in good approximation, the dominant wavelength only depends 
on the relation $L/w$ independently of the remaining parameters.  
However, these parameters  participate in conditioning pattern upraise  
through the critical value of $\beta$, 
given by Eq.~(\ref{cond2}), and may also influence pattern shape.    
In particular, in the average, $m$ does not depend on $\nu$, as  can  be observed in the  example 
of Fig.~\ref{fig:patterns}. 
The wavenumber is preserved in the limit $D\to0$, even if in 
approaching this limit more and more modes become unstable (i.e., more 
local maxima of $\Lambda_k$ become positive), but the global maximum remains approximately the same. 
Then, diffusion is not necessary  for  
pattern formation, nor influences the characteristic wavelength. 

\begin{figure} [h!]
\centering
\includegraphics*[bb=70 460 500 750, width=0.95\columnwidth]{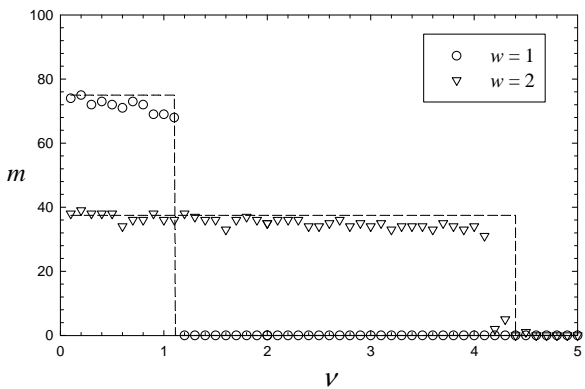}
\caption {Number of maxima $m$ as a function of $\nu$ for two different values of 
$w$ indicated on the figure, with  
$a=b=1$, $L=100$ and $D=0.01$. 
Dashed lines correspond to the theoretical prediction given by Eq. (\ref{aprox}), 
with the additional condition (\ref{cond2}), defining a critical value $\nu_c$ beyond which 
no patterns occur (we set  $m=0$ in such case).  
} 
\label{fig:critical}
\end{figure}

Notice also that condition (\ref{cond2}) indicates that, although  
approximately the same number of maxima is expected  independently of $\nu$, 
there are critical thresholds $\nu_c$ beyond which no patterns occur. 
This is illustrated in Fig. \ref{fig:critical} for the case $a=b=1$, 
for which $\nu_c= w^2/( D\theta_0^3)$.

\begin{figure} [t!]
\centering    
\includegraphics*[bb=70 460 500 750, width=0.95\columnwidth]{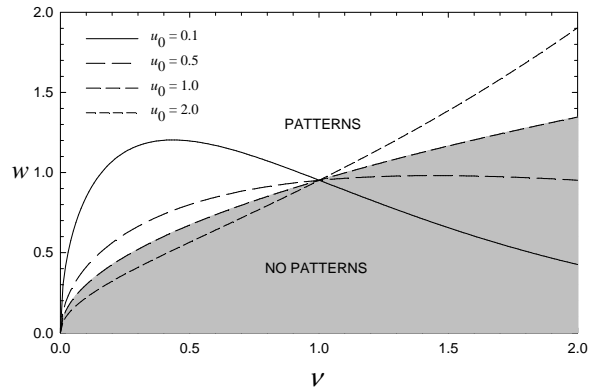}
\caption {Phase diagram  of pattern formation in the $\nu\,w$ plane , 
for $u_0=1$, following Eq.~(\ref{cond2}). 
Shadowed is the region where no patterns arise. 
The lines show the frontier of the phase diagram for other values of 
$u_0=a/b$. Patterns emerge for $w>w_c$ (above the critical line). 
Parameter values are $L=100$, $D=0.01$ and  $a=1$. 
} 
\label{fig:diagram}
\end{figure}

Differently from normal diffusion, $u_0=a/b$ 
is also determinant of pattern formation. 
How the phase diagram is altered by changes in $u_0$ is illustrated in Fig. \ref{fig:diagram}. 
The shadowed area represents the region where no patters emerge when $u_0=1$. 
For other values of $u_0$, only the frontier is shown. 
For $u_0\ge 1$, the critical curve increases monotonically with $\nu$, 
such that only small $\nu$ (superdiffusion) allows
pattern formation for a given interaction range $w$ and remaining parameters fixed. 
However,  the monotonic behavior of the critical curve is broken when $u_0<1$. 
Thus,  for low values of $w$ there is also an upper critical value of $\nu$ for the onset 
of patterns, but for large $w$, patterns occur for any $\nu$.

\subsection{Patterns shape}

\begin{figure} [b!]
\centering
\includegraphics*[bb=70 450 500 750, width=0.95\columnwidth]{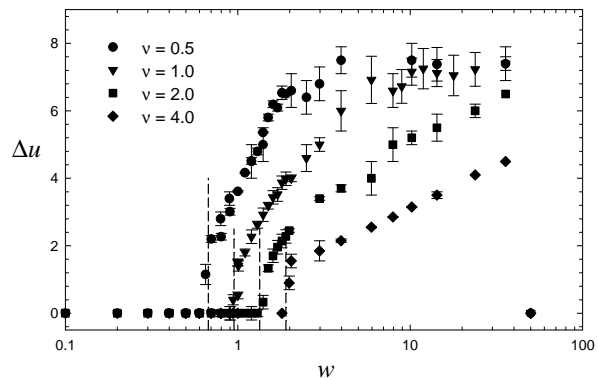}
\caption {Pattern amplitude $\Delta u\equiv u_{max}-u_{min}$ 
as a function of the interaction width $w$, for 
$a=b=1$, $L=100$, $D=0.01$, and different values of $\nu$ indicated on the figure.
The dotted lines are a guide to the eyes, the vertical ones indicate the critical 
value predicted by Eq. (\ref{cond2}). 
} 
\label{fig:A}
\end{figure}

Although the characteristic mode does not  depend on $\nu$, its amplitude does.  
This is shown in Fig.~\ref{fig:A} where 
the amplitude, $\Delta u=u_{max}-u_{min}$ obtained 
from numerical simulations is represented as a function of $w$ for different values of $\nu$.  
 
For vanishing $w$ we recover the local case $f(x,x') = \delta(x-x')$ 
in which no patterns emerge, as supported by numerical simulations. 
In agreement with Eq.~(\ref{cond2}), there is a critical value $w_c$, 
at which the amplitude vanishes. 
Notice the abrupt decay of the amplitude at the critical value.
This threshold was not detected in previous works dealing with normal diffusion 
possibly because of the range 
of parameters used. For instance in Ref.~\cite{fuentes}, $w_c/L$ would be of the 
order of $10^{-3}$.
The critical value $w_c$ decreases with $\nu$, 
indicating that a shorter influence range $w$ is required when the dispersion passes 
from subdiffusive to superdiffusive.
Then superdiffusion favors pattern formation and the amplitude of the patterns is larger.
For $w=L/2$ (or its multiples), the nonlocal term becomes $J[u(x,t)]=J(t)$, 
that follows the equation $dJ/dt=(a-bJ)J$. Then, in the long time limit $J \to u_0$, implying that 
the homogeneous state should also be attained in this extreme case.

Furthermore,  $\nu$ can have strong effects on patterns shape.  
As depicted in Fig.~\ref{fig:patterns}, 
subdiffusion ($\nu>1$) induces fragmentation (clusterization). 
Solutions that vanish outside a finite support are 
typical of nonlinear subdiffusion \cite{murray}.
In the opposite case   $\nu<1$ (super-diffusion), the effects are not so striking 
concerning pattern shape, for moderate values of the diffusion coefficient. 
The region between crests assumes larger values the smaller $\nu$.  
Fragmentation also emerges  for any kind of diffusion when $D$ is small enough 
(as discussed in connection to Fig.~\ref{fig:zerodif}) or also if $w$ becomes large enough 
(not shown).
The shape of the clusters depends on $\nu$. Their  amplitude decays and their width increases as 
$\nu$ increases.

It is noteworthy that the distance between crests (wavelength),  
$\ell = L/m\simeq 1.4w$,  is   larger than the interaction range $w$,  
however if the cluster size $2x_0(w)$  is large enough, 
there can be influence of one cluster over the two neighboring ones. 
When clusters are disconnected, $2x_0 \lesssim \ell-w= 0.4 w$ means that one cluster 
does not influence the neighbors. Otherwise they do, even if disconnected.

\section{Final remarks}

Nonlinear diffusion is expected in the spreading of biological populations 
rather than normal diffusion, hence motivating the introduction of a state 
dependent diffusion coefficient, as in Eq.~(\ref{gfisher}). 
We have shown  how pattern formation is altered in the presence of anomalous diffusion. 
Moreover, in all cases, the initially fastest growing mode remains selected at longer times. 
This observation allows 
to obtain theoretical predictions that we verified through numerical integration 
of the evolution equation.  

Then, it is clear that diffusion is not a necessary ingredient for the onset of patterns, 
nor has impact on the characteristic wavelength, 
that depends only on the interaction range $w$.  
Furthermore, diffusion imposes a critical threshold of the model parameters for pattern formation. 
The type of diffusion regime has impact on patterns shape, even  if 
the characteristic mode is kept unchanged. 
An important qualitative change in the shape of 
patterns occurs mainly for $\nu>1$, in which case fragmentation of the population 
is induced. This effect is also observed for very small diffusion constant $D$ and/or large 
interaction width $w$. 
The occurrence of fragmentation may have important consequences in 
diseases dissemination and other spreading processes triggered 
by contacts between individuals.  
Superdiffusion ($\nu<1$) facilitates pattern formation, which can occur even for shorter 
interaction width  $w$   and manifesting larger amplitudes 
 than in normal diffusion.

Beyond the initial motivation of introducing nonlinear diffusion to the nonlocal FE, 
we uncovered aspects that apply also to the normal 
diffusion case previously studied. The identification of the main mode selected 
at long times  allows to perform  analytical predictions, which may be extended 
 to tackle other variants of the model.

\begin{acknowledgments}

We are grateful to Welles A.M. Morgado for useful discussions. 
We acknowledge partial financial support from CNPq and Capes  
(Brazilian Government Agencies).

\end{acknowledgments}

\end{document}